\documentclass[12pt,a4paper]{article}
\usepackage{amsfonts}

\begin{document}

\title{Non-chiral fusion rules and structure constants of $D_{m}$ minimal models.}
\author{A. Rida \\
{\small Laboratoire de physique th\'{e}orique, Institut de Physique,}\\
{\small Universit\'{e} des Sciences et de la Technologie Houari-}\\
{\small Boumedienne, BP.32, El Alia Alger, 16111 Algeria.} \and T. Sami%
\thanks{%
e-mail: Sami@subatech.in2p3.fr } \\
{\small Laboratoire SUBATECH, Ecole des Mines de Nantes; \ 4, \ Rue }\\
{\small Alfred Kastler, La Chantrerie 44070 Nantes, France.}}
\date{}
\maketitle

\begin{abstract}
We present a technique to construct, for $D_{m}$ unitary minimal models, the
non-chiral fusion rules which determines the operator content of the
operator product algebra. Using these rules we solve the bootstrap equations
and therefore determine the structure constants of these models. Through
this approach we emphasize the role played by some discrete symmetries in
the classification of minimal models.

\medskip
\end{abstract}

To be submitted to Int. J. Mod. Phys. A.

\section{Introduction}

The first and the most known process to deal with the classification problem
of two dimensional conformal field theories (CFT) is the bootstrap approach.
This approach initially developed in the seminal work of Belavin, Polyakov
and Zamolodchikov (PBZ) \cite{BPZ} is based, among other things, on the
associativity property of the operator product algebra (OPA) of the four
point correlation functions on the plane. Formally speaking, this property
known as crossing symmetry is expressed by the so-called bootstrap
equations. These equations are the master equations since their resolution
gives in principle a complete classification of a conformal theory. (PBZ)
mentioned the existence of a particular class of (CFT) associated to
degenerate representation of the Virasoro algebra for which the bootstrap
equations can be solved. This class of CFT's designed as minimal models
include the finite discrete unitary models with central charge $c<1$ \cite
{FQS}. \ 

In this framework, Dotsenko and Fattev (DF) \cite{Dot} proposed to solve the
bootstrap equations and therefore they determined the structure constants of
the operator algebra \ for correlations with spinless fields. They use in
their construction the monodromy property of conformal blocs defining the
coordinate dependencies of the correlation functions in the halomorphic
(antihalomorphic) sector $z\left( \overline{z}\right) $ in the Coulomb gas
formalism. Indeed for spineless fields the conformal blocs in the two
sectors are complex conjugate so that only diagonal combinations of this
blocs survive to the monodromy constraint. After this work Capelli, Itzykson
and Zuber \cite{CIZ} made a complete classification of modular invariant
partition functions: the $ADE$ classification of minimal conformal models.
Their results gave the operator content of all the minimal models. The $%
\left( A\right) $ present only spinless primary fields and the models of
this series are those studied by (DF).

Nevertheless, in the case of the $\left( D\right) $ series solving the
bootstrap equations seems to be more difficult. Two difficulties arise;
namely the presence of spin fields and the existence of two copies of
certain fields. Because of the presence of the non-vanishing spin fields, it
is difficult to formulate the monodromy constraint in the (DF) coulomb gas
approach spirit. Moreover, with the monodromy constraint alone we cannot
differentiate the behavior of the different copies of doubled fields in the
(OPA).

Many approaches have been developed to solve these problems. Apparently the
most complete and general approach is the one developed by Petkova \cite
{Petko}. This consists globally in an adaptation of the principle of
monodromy in the (DF) coulomb gas formalism for the $\left( D\right) $
series models. The importance of this adaptation resides in that it permits
to understand the technical aspects of the (DF) formulation. However,
despite its importance, which will be of a great utility for us, this
approach present many ambiguities regarding its principle \cite{John}%
\footnote{%
One of these adaptations is the normalization of the two point correlations
by a factor equal to $\left( -1\right) ^{s}$ rather than $1$, where $s$ is
the spin two point field. This redefinition of the normalization factor is
in complete contradiction with unitarity. In fact, the inner product $I$ of
two of the highest weight states is determined by \cite{BPZ} \cite{Moore}: 
\begin{eqnarray*}
I\left( \left| \alpha \right\rangle ,\left| \beta \right\rangle \right)
=\lim_{z\rightarrow \infty ,\,\,w\rightarrow 0}z^{2h}\overline{z}^{2%
\overline{h}}\left\langle 0\right| \Phi ^{\alpha }\left( z,\overline{z}%
\right) \Phi ^{\beta }\left( w,\overline{w}\right) \left| 0\right\rangle \\
=\left( -1\right) ^{s_{\alpha }}\delta _{\alpha \beta }
\end{eqnarray*}
\par
We see that the factor $\left( -\right) ^{s}$ would give negative norm
states.}. It is important to mention the existence of another approach \cite
{Zuber} which call on other formalisms namely the lattice representation (
generalized RSOS models) of the $ADE$ models \cite{RSOS}. In this approach
the ratios of the structure constants of the $D$ or $E$ theories over the
corresponding structure constants of the $A$ theory of the same Coxeter
number are determined. It was also mentioned in Ref.\cite{Zuber} that these
results \textit{do not seem easy derived from crossing (bootstrap) equations}
and only \textit{phenomenological} observations are given.

What we propose in the present work is a simple and general approach to
solve the bootstrap equations without using the monodromy constraint. Our
approach is a generalization of the ideas initially developed in Ref.\cite
{John1} for the particular case of the $D_{5}$ $\left( c=\frac{3}{5}\right) $
model. The principle of the idea lays on the construction of the so called 
\emph{non-chiral fusion rules.} The non-chiral fusion rules determine the
operator content of the operator algebra. Initially the term fusion rules
was used to express how two representations of the Virasoro algebra combine
in the (OPA). The non-chiral fusion rules permit to avoid the monodromy
problem and hence to make a new step in the resolution of the bootstrap
equations. For the determination of the non-chiral fusion rules we use some
considerations imposing strict constraints. The first constraint lays on the
consistency of the non-chiral fusion rules and the fusion rules (chiral).
The second consideration, rather obvious, consist on imposing the
consistency of the non-chiral fusion rules with the operator content
determined by the modular constraint. The third important consideration,
which is given to construct the non-chiral fusion rules lays on symmetry
considerations. This last consideration is manifested by a discrete symmetry 
$Z_{2}$ which is defined by its action on the two components of a doubled
field $\left( \Phi ^{\pm }\right) $: $Z_{2}\left( \Phi ^{\pm }\right) =\pm
\Phi ^{\pm }.$ Thus, this symmetry $\left( Z_{2}\right) $ permit to separate
the doubled field contribution by imposing the consistency of the non-chiral
fusion rules with its action. Under the action of the $Z_{2}$ symmetry the
scalar fields are found to have a positive parity contrary to spin fields
which have negative parity. As a consequence; the consistency of the
non-chiral fusion rules with the $Z_{2}$ action gives them a $Z_{2}-grading$
structure.

Once the non-chiral fusion rules are determined, it is possible to solve the
bootstrap equations by considering that at short distance these equations
should be consistent with the (OPA) which we have expressed through these
rules. This is precisely what was done in Ref.\cite{John1} for the $D_{5}$
model. Here, we propose a generalization of these calculations to the all
set of $D_{m}$ models. For \ this end we use general analytic and duality
proprieties of conformal blocs in the coulomb gas formalism developed in
Ref. \cite{Petko}. As a result, we have found that the structure constants
of the $D_{m}$ series models factories out in those of the chiral algebra
expressed by the $A_{m}$ series. The signs of the structure constants are
also determined.

In our construction a particular role is played by $Z_{2}$ symmetry which
reflects the $Z_{2}-grading$ structure of the non-chiral fusion rules. A
physical interpretation of the $Z_{2}$ symmetry as a scaling limit symmetry
of some lattice models namely the generalized RSOS \cite{RSOS} models is
given. Furthermore, we show the importance of discrete symmetries and
particularly the symmetry $Z_{2}$ in the classification of minimal models.
These remarks make our approach applicable in general for the large set of
rational conformal field theories \cite{Amar}.

This article is organized as follows. In sec.2 we briefly review some
important results of minimal models classification. We give in particular a
precise analysis of the operator content spectrum of $\left( D_{m}\right) $
unitary series. In sec.3 we present our approach by constructing the
non-chiral fusion rules. In sec.4 we solve the bootstrap equations and
determine the structure constants. In sec.5 we give a general analysis of
our approach and extract some important consequences and we conclude in
sec.6.

\section{General features of minimal models}

One of the fundamental requirements in conformal field theories is the
existence of a closed operator algebra. At two dimensions this requirement
is expressed by the operator product algebra of two primary fields which can
be written in the form: 
\begin{equation}
\Phi _{\mathbf{I}}\left( z,\overline{z}\right) \cdot \Phi _{\mathbf{J}%
}\left( w,\overline{w}\right) =\sum_{\mathbf{K}}C_{\mathbf{IJK}}\left(
z-w\right) ^{h_{k}-h_{i}-h_{j}}\left( \overline{z}-\overline{w}\right) ^{%
\overline{h}_{k}-\overline{h}_{i}-\overline{h}_{j}}\left[ \Phi _{\mathbf{K}%
}\left( w,\overline{w}\right) +\ldots \right] .  \label{OPA}
\end{equation}
where $I=\left( i,\overline{i}\right) $ indicates different fields with $%
i\left( \overline{i}\right) $ representing the chiral contributions of the
halomorphic (antihalomorphic) sectors. $h_{i}$ and $\overline{h}_{i}$ are
the conformal dimensions of $\Phi _{\mathbf{I}}$ and the ellipsis stands for
terms involving descendant fields. $C_{\mathbf{IJK}}$ are complex numbers
known as structure constants of the operator algebra.

The associativity of the operator product algebra in four point correlation
functions on the plane yields to symmetry relations termed as bootstrap
equations: 
\begin{eqnarray}
G_{\mathbf{IJKP}}\left( z,\overline{z}\right) &=&\left\langle \Phi _{\mathbf{%
I}}\left( 0\right) \cdot \Phi _{\mathbf{J}}\left( z,\overline{z}\right)
\cdot \Phi _{\mathbf{L}}\left( 1\right) \cdot \Phi _{\mathbf{K}}\left(
\infty \right) \right\rangle  \label{G} \\
&=&\sum_{\left( p,\overline{p}\right) }C_{\mathbf{IJP}}C_{\mathbf{LMP}}%
\mathcal{F}_{ij}^{lm}\left( p\mid z\right) \overline{\mathcal{F}}_{\overline{%
i}\,\overline{j}}^{\overline{l}\,\overline{m}}\left( \overline{p}\mid 
\overline{z}\right)  \label{1-s} \\
\ \, &=&\sum_{\left( q,\overline{q}\right) }C_{\mathbf{ILQ}}C_{\mathbf{JMQ}}%
\mathcal{F}_{lj}^{im}\left( q\mid 1-z\right) \overline{\mathcal{F}}_{%
\overline{l}\,\overline{j}}^{\overline{i}\,\overline{m}}\left( \overline{q}%
\mid 1-\overline{z}\right) .  \label{2-t}
\end{eqnarray}
{\ }where $\mathcal{F}_{ij}^{lm}\left( p\mid z\right) $ denotes the
conformal blocs. These two expressions (\ref{1-s}) and (\ref{2-t}) of the
correlation $G$ are noted as t-channel and s-channel development
respectively.

In the case of degenerate representations of the Virasoro algebra projecting
out null states imposes a strict constraint on conformal models. This yields
a particular class of conformal theories known as minimal models. For
unitary minimal models the central charge and the conformal dimensions of
fields are confined to discrete values: 
\begin{eqnarray*}
c\left( m\right) &=&1-\frac{6}{m\left( m+1\right) }, \\
h_{rs}\left( m\right) &=&h_{m-r,m+1-s}\left( m\right) =\frac{\left( \left(
m+1\right) r-ms\right) ^{2}-1}{4\left( m+1\right) m},\, \\
\,1 &\leq &r\leq m-1,\,\,1\leq s\leq m.
\end{eqnarray*}
In these cases the conformal blocs are solutions of differential equations.
This later important fact, proved to give hard constraints on the operator
algebra. These constraints are expressed through the fusion rules, which
determine how the chiral parts of physical fields combine in the (OPA). If
we associate to each primary field $\Phi _{\mathbf{I}}$, $\mathbf{I=}$ $%
\left( r,s\mid \overline{r},\overline{s}\right) $, its chiral part $\left(
r,s\right) $; the fusion rules for minimal models will be given by:{\ 
\begin{equation}
\left( r_{1},s_{1}\right) \times \left( r_{2},s_{2}\right) =\sum_{k=\left|
r_{1}-r_{2}\right| +1}^{\min \left( r_{1}+r_{2},2m-r_{1}-r_{2}+1\right)
}\,\,\,\,\,\,\,\,\,\,\sum_{l=\left| s_{1}-s_{2}\right| +1}^{\min \left(
s_{1}+s_{2},2m-s_{1}-s_{2}+1\right) }\left( k,l\right) .  \label{fusion}
\end{equation}
}

{On another hand, }the modular constraint on the partition function, give a
complete classification of the operator content of minimal models: $ADE$
patterns \cite{CIZ}. First the $\left( A\right) $ series is diagonal so that
the operator content is composed by spinless scalar fields. For the $D$
series the partition functions are non-diagonal. They can be organized in
the unitary case as follows:

\begin{itemize}
\item  $m=4\rho +2$%
\begin{equation}
Z=\frac{1}{4}\sum_{s=1}^{m}\sum_{r(odd)=1}^{m-1}\left| \chi _{rs}+\chi
_{m-r,s}\right| ^{2}.  \label{1}
\end{equation}

\item  $m=4\rho +1$%
\begin{equation}
Z=\frac{1}{4}\sum_{r=1}^{m-1}\sum_{s(odd)=1}^{m}\left| \chi _{rs}+\chi
_{r,m+1-s}\right| ^{2}.  \label{2}
\end{equation}

\item  $m=4\rho $%
\begin{equation}
Z=\frac{1}{2}\sum_{s=1}^{m}\left\{ \sum_{r(odd)=1}^{m-1}\left| \chi
_{rs}\right| ^{2}+\sum_{r(even)=2}^{m-2}\chi _{m-r,s}\chi _{rs}^{\ast
}\right\} .  \label{3}
\end{equation}

\item  $m=4\rho +3$%
\begin{equation}
Z=\frac{1}{2}\sum_{r=1}^{m-1}\left\{ \sum_{s(odd)=1}^{m}\left| \chi
_{rs}\right| ^{2}+\sum_{s(even)=2}^{m-1}\chi _{r,m+1-s}\chi _{rs}^{\ast
}\right\} .  \label{4}
\end{equation}
\end{itemize}

\subsection{Spectrum analysis}

From the partition functions of the $D$ series unitary minimal models
presented above, we can summarize the operator content spectrum with respect
to $m$ values modulo $4$ as follows:

\begin{center}
\begin{tabular}{cccc}
$m$ & scalar fields & spin fields &  \\ 
\multicolumn{1}{l}{$m=2%
\bmod%
%
4$} & \multicolumn{1}{l}{$\left( r,s\mid r,s\right) $} & \multicolumn{1}{l}{$%
\left( r,s\mid m-r,s\right) $} & \multicolumn{1}{l}{$r=odd$} \\ 
\multicolumn{1}{l}{$m=1%
\bmod%
%
4$} & \multicolumn{1}{l}{$\left( r,s\mid r,s\right) $} & \multicolumn{1}{l}{$%
\left( r,s\mid r,m+1-s\right) $} & \multicolumn{1}{l}{$s=odd$} \\ 
\multicolumn{1}{l}{$m=0%
\bmod%
%
4$} & \multicolumn{1}{l}{$\left( r,s\mid r,s\right) \left( r=odd\right) $} & 
\multicolumn{1}{l}{$\left( r,s\mid m-r,s\right) \left( r=even\right) $} & 
\multicolumn{1}{l}{} \\ 
\multicolumn{1}{l}{$m=3%
\bmod%
%
4$} & \multicolumn{1}{l}{$\left( r,s\mid r,s\right) \left( s=odd\right) $} & 
\multicolumn{1}{l}{$\left( r,s\mid r,m+1-s\right) \left( s=paire\right) $} & 
\multicolumn{1}{l}{}
\end{tabular}
\end{center}

\smallskip

\noindent According to the operator spectrum one can distinguish two
categories:

\begin{itemize}
\item  $m=0%
\bmod%
%
4$ and $m=3%
\bmod%
%
4:$ these cases are noted as automorphism or permutation invariant
solutions. This notation comes from the fact that the modular invariant
partition functions can be written as: $Z=\sum_{\left( rs\right) }\chi
_{rs}\chi _{\mu \left( rs\right) }$, where $\left( \mu \right) $ is some
automorphism of fusion rules \cite{Auto}. We remark in this category the
existence of a difference in parity with respect to the indices $r$ or $s$
between the scalar and spin fields. Also, for the values $r=m/2$ ($m=0%
\bmod%
%
4$) and $s=\left( m+1\right) /2$ ($m=3%
\bmod%
%
4$) the corresponding spin fields have a null spin; they are thus scalar
fields. For technical reasons, that we will see later, they are called null
spin field.

\item  $m=2%
\bmod%
%
4$ and $m=1%
\bmod%
%
4:$ these cases are noted as integer invariant solutions. The partition
functions of this category (\ref{1} - \ref{2}) can be regarded as diagonal
invariant solutions of a larger chiral algebra than originally considered
(Virasoro algebra) \cite{Moore}. The field extending this algebra is
represented by a character appearing in the same term as the identity
character. This field is thus: $\left( 1,1\mid 1,\frac{m+1}{2}\right) $ and $%
\left( 1,1\mid 1,\frac{m}{2}\right) $ for $m=2%
\bmod%
%
4$ and $m=1%
\bmod%
%
4$ respectively. What characterizes the operator content of these cases in
not the difference in parity between the scalar and spin fields with respect
to the indices $\left( r,s\right) $, but \ the appearance of two copies of
certain scalar fields. In fact, for $m=2%
\bmod%
%
4$ $\left( m=1%
\bmod%
%
4\right) $ with $r=m/2$ $\left( s=\left( m+1\right) /2\right) $, the
corresponding spin fields have the forme: 
\begin{eqnarray}
\left( m/2,s\mid m/2,s\right) ,\,\,\,\,\,m=2%
\bmod%
%
4.  \label{copie1} \\
\left( r,\left( m+1\right) /2\mid r,\left( m+1\right) /2\right)
,\,\,\,\,\,m=1%
\bmod%
%
4.  \label{copie2}
\end{eqnarray}
and have, thus, a forme of a null spin fields. Furthermore we note that the
same fields are present in the scalar fields set so that we have two copies
of these fields\linebreak (\ref{copie1} - \ref{copie2}) in the operator
content spectrum. We note these two copies by $\ \Phi ^{+}$ and $\Phi ^{-}$.
\end{itemize}

The operator content analysis in the case of the $D$ unitary minimal models
we have developed will permit us later to give a simple and general
description of the operator algebra through the non-chiral fusion rules. At
this point it is interesting to note that the two cases $m=2%
\bmod%
%
4$ and $m=1%
\bmod%
%
4$ are symmetric via the permutations $s\leftrightarrow r$, $%
m\leftrightarrow m+1.$ This same remark is valid for $m=0%
\bmod%
%
4$ and $m=3%
\bmod%
%
4.$ Thanks to this, the treatment of the $D$ series unitary models can be
reduced to the cases $m=3%
\bmod%
%
4$ for the automorphism invariant cases and $m=1%
\bmod%
%
4$ for the integer invariant cases.

\section{Non-chiral fusion rules}

By definition the non-chiral fusion rules determine the operator content of
the operator algebra. Therefore, \emph{the fusion of the two fields }$\Phi _{%
\mathbf{I}}$\emph{\ and }$\Phi _{\mathbf{J}}$\emph{\ produce the field }$%
\Phi _{\mathbf{K}}$\emph{\ if and only if the structure constant }$C_{%
\mathbf{IJK}}$\emph{\ is non-vanishing}. To construct the non-chiral fusion
rules in the case of minimal models we will use two important facts which
are the consistency of these rules with the fusion rules (chiral) and the
operator content structure together

\noindent In fact, from the bootstrap equations we can get the following
condition: 
\begin{equation}
C_{\mathbf{IJK}}\neq 0\Rightarrow \left\{ 
\begin{array}{c}
\left( i\right) \times \left( j\right) \rightarrow \left( k\right) \\ 
\left( \overline{i}\right) \times \left( \overline{j}\right) \rightarrow
\left( \overline{k}\right)
\end{array}
\right. .  \label{base1}
\end{equation}
with $\left( i\right) \times \left( j\right) \rightarrow \left( k\right) $
translating the fusion condition of $\left( i\right) $ and $\left( j\right) $
which give the field $\left( k\right) .$ This condition indicates the
consistency of the non-chiral fusion rules with the fusion rules. The second
condition that must be considered is the compatibility of the operator
algebra with the operator content spectrum. If we designate this spectrum by
the set $\mathcal{A}$ this condition becomes: 
\begin{equation}
C_{\mathbf{IJK}}\neq 0\Rightarrow \mathbf{I,J,K\in }\mathcal{A}.
\label{base2}
\end{equation}

For the series $\left( A\right) $ models the application of these conditions
lead to the known result obtained by (DF) by the monodromy invariance \cite
{Dot}: 
\begin{equation}
\left( s_{1},r_{1}\mid s_{1},r_{1}\right) \times \left( s_{2},r_{2}\mid
s_{2},r_{2}\right) =\sum_{k=\left| r_{1}-r_{2}\right| +1}^{\min \left(
r_{1}+r_{2},2m-r_{1}-r_{2}+1\right) }\,\,\,\,\,\,\,\,\,\,\sum_{l=\left|
s_{1}-s_{2}\right| +1}^{\min \left( s_{1}+s_{2},2m-s_{1}-s_{2}+1\right)
}\left( k,l\mid k,l\right) .  \label{FusionA}
\end{equation}

\subsection{$D$ series non-chiral fusion rules}

\subsubsection{Automorphism invariant cases}

Due to the different parities of indices indicating the scalar and spin
fields, the non-chiral fusion rules can be easily determined.

\vspace{0.5cm}      %
%

\noindent \textbf{Example of }$D_{7}$ \textbf{model}: For this model the
partition function (\ref{4}) produce the following operator content: 
\begin{eqnarray}
Z &=&\chi _{11}\chi _{11}^{\ast }+\chi _{13}\chi _{13}^{\ast }+\chi
_{15}\chi _{15}^{\ast }+\chi _{17}\chi _{17}^{\ast }+\chi _{16}\chi
_{12}^{\ast }+\chi _{14}\chi _{14}^{\ast }+\chi _{12}\chi _{16}^{\ast } 
\nonumber \\
&&+\chi _{51}\chi _{51}^{\ast }+\chi _{53}\chi _{53}^{\ast }+\chi _{55}\chi
_{55}^{\ast }+\chi _{57}\chi _{57}^{\ast }+\chi _{56}\chi _{52}^{\ast }+\chi
_{54}\chi _{54}^{\ast }+\chi _{52}\chi _{56}^{\ast }  \nonumber \\
&&+\chi _{31}\chi _{31}^{\ast }+\chi _{33}\chi _{33}^{\ast }+\chi _{35}\chi
_{35}^{\ast }+\chi _{37}\chi _{37}^{\ast }+\chi _{36}\chi _{32}^{\ast }+\chi
_{34}\chi _{34}^{\ast }+\chi _{32}\chi _{36}^{\ast }.  \label{exeD7}
\end{eqnarray}
We begin by determining the fusion rules (\ref{fusion}) of the model. Among
other things we find: 
\begin{eqnarray*}
\left( 15\right) \times \left( 15\right) &=&\left( 11\right) +\left(
13\right) +\left( 15\right) , \\
\left( 17\right) \times \left( 53\right) &=&\left( 55\right) , \\
\left( 16\right) \times \left( 14\right) &=&\left( 13\right) +\left(
15\right) ,\,\,\,\,\,\,\,\,\,\,\left( 12\right) \times \left( 16\right)
=\left( 15\right) +\left( 17\right) \,\, \\
\left( 16\right) \times \left( 36\right) &=&\left( 31\right) +\left(
33\right) ,\,\,\,\,\,\,\,\,\,\,\left( 12\right) \times \left( 32\right)
=\left( 31\right) +\left( 33\right) \\
\left( 15\right) \times \left( 12\right) &=&\left( 14\right) +\left(
16\right) ,\,\,\,\,\,\,\,\,\,\,\left( 15\right) \times \left( 15\right)
=\left( 12\right) +\left( 14\right) .
\end{eqnarray*}
Finally to deduce the non-chiral fusion rules of the model $D_{7}$, we
combine the fusion rules in the two sectors and use the consistency with the
operator content. As a final result we obtain:

\vspace{0.3cm}      %
%

\begin{center}
\begin{tabular}{lr}
$\left( 1,5\mid 1,5\right) \times \left( 1,5\mid 1,5\right) =\left( 1,1\mid
1,1\right) +\left( 1,3\mid 1,3\right) +\left( 1,5\mid 1,5\right) $ & $\left(
a\right) $ \\ 
$\left( 1,7\mid 1,7\right) \times \left( 5,3\mid 5,3\right) =\left( 5,5\mid
5,5\right) \,\,\,\,\,\,$ & $\,\,\,\left( b\right) $ \\ 
$\left( 1,6\mid 1,2\right) \times \left( 1,4\mid 1,4\right) =\left( 1,5\mid
1,5\right) \,\,\,\,\,\,\,\,\,\,$ & $\,\,\left( c\right) $ \\ 
$\left( 1,5\mid 1,5\right) \times \left( 1,6\mid 1,2\right) =\left( 1,4\mid
1,4\right) +\left( 1,2\mid 1,6\right) \,$ & $\left( d\right) $ \\ 
$\left( 1,6\mid 1,2\right) \times \left( 3,6\mid 3,2\right) =\left( 3,3\mid
3,3\right) +\left( 3,1\mid 3,1\right) $ & $\left( e\right) $%
\end{tabular}
\end{center}

\vspace{0.3cm}      %
%

We remark that the fusion of two scalar fields or two spin fields produce
only scalar fields (\textrm{rule }$\left( a\right) $\textrm{, }$\left(
b\right) $\textrm{\ }and\textrm{\ }$\left( e\right) $). Whereas, the fusion
of scalar fields and spin fields produce only spin fields (\textrm{rule }$%
\left( d\right) $). It is to be noted that the field $\left( 1,4\mid
1,4\right) $ \emph{which is a null spin field }behaves like a spin field in
the non-chiral fusion rules (\textrm{rule }$\left( c\right) $). This
justifies the name null spin field.

The conclusions of the preceding example are valid in the most general case.
This is due, indeed, to the difference in parity of the indices between the
scalar fields and spin fields. In fact, for $m=3%
\bmod%
%
4$ the index $s$ is odd for scalar fields and even for spin fields. As a
consequence, the fusion products have an index $s$ with the parity of $%
\left| s_{1}-s_{2}\right| +1$, that is: odd if two fields have the same
parity and even if they have different parity. By symmetry $s\leftrightarrow
r$, the same argument is valid for the case $m=0%
\bmod%
%
4$.

In conclusion, if we designate by $A_{0}$ the set of scalar fields and $%
A_{1} $ the set of spin and null spin fields, the non-chiral fusion rules
will have the forme: 
\[
A_{i}*A_{j}=A_{k},\,\,\,k=\left( i+j\right) 
\bmod%
%
2 
\]
The so constructed non-chiral fusion rules have thus a $``Z_{2}-grading"$
structure.

These fusion rules structure we have just determined translates an important
fact (especially for the remaining) which is the conservation of a parity in
these rules. Indeed, if we affect a positive parity charge to the set of
scalar fields $\left( A_{0}\right) $ and a negative parity charge to the set
of spin and null spin fields $\left( A_{1}\right) $ then the $Z_{2}-$grading
structure translates the conservation of this charge in the non-chiral
fusion rules.

\subsection{Integer invariant cases}

For these $D_{m}$ series models there is no difference in parity between the
scalar and spin fields and it is not possible to deduce simple rules as
above. In addition and as we have already remarked there is a doubling of
certain fields. The two components of these fields which we have noted $\Phi
^{+}$ and $\Phi ^{-}$ will have the same contribution in the (OPA) and it is
not possible, thus, to distinguish between the behavior of each one in a
correlation function.

To overcome these difficulties we use an important physical fact namely that
by the state-field correspondence principle the presence of many copies of
primary fields translates the degeneracy of the ground state. To lift this
degeneracy we introduce a discrete parity symmetry $Z_{2}$: 
\begin{equation}
Z_{2}\left( \Phi ^{\pm }\right) =\pm \Phi ^{\pm }
\end{equation}
The characterization of the $Z_{2}$ symmetry is complete if one arrives at
defining its action on the other fields of the model $\left\{ \Phi _{\alpha
}\right\} $. To this end we use an important consideration which is the
consistency of the operator product algebra with the action of this
symmetry. This last consideration is expressed by adding to the two
construction bases (\ref{base1} - \ref{base2}) a third constraint which is
the consistency of the non-chiral fusion rules with the $Z_{2}$ symmetry
action, i.e. the different members in a fusion rule must have the same
parity (the conservation of $Z_{2}$ parity charge in the non-chiral fusion
rules).

For the $Z_{2}$ symmetry construction we use the $D_{m}$ series with $m=1%
\bmod%
%
4.$ The same argument is valid by symmetry for $m=2%
\bmod%
%
4.$

\subsubsection{Construction of $Z_{2}$}

The technique of a $Z_{2}$ parity symmetry in the construction of the
non-chiral fusion rules was initiated in Ref.\cite{John1} for the particular
case of the $D_{5}$ model. In this work, the authors were concerned only
with thermic sub-algebra: 
\[
\left\{ \left( 15,15\right) ,\left( 15,11\right) ,\left( 11,15\right) ,\Phi
_{13}^{+},\Phi _{13}^{-}\right\} . 
\]
The construction of this symmetry is based on the consistence of its action
with the operator algebra. Thus, from the fact that: 
\[
\left( 15,15\right) \times \left( 15,11\right) =\left( 11,15\right) 
\]
one can see that the $Z_{2}$ action will be limited to the following cases 
\cite{John1}:

\pagebreak

\begin{enumerate}
\item[A)]  {\small $\,\left( 15;15\right) $ $\rightarrow +\left(
15;15\right) ,$ $\,\,\,\,\,\left( 11;15\right) \rightarrow +\left(
11;15\right) ,\,\,\,\,\,\,\,\left( 15;11\right) \rightarrow +$ $\left(
15;11\right) $ \ }

\vspace{0.25cm}

\item[B)]  {\small $\,\left( 15;15\right) $ $\rightarrow +\left(
15;15\right) ,$ $\,\,\,\,\,\left( 11;15\right) \rightarrow -\left(
11;15\right) ,\,\,\,\,\,\,\,\left( 15;11\right) \rightarrow -$$\left(
15;11\right) $ \ }

\vspace{0.25cm}

\item[C)]  {\small $\,\left( 15;15\right) $ $\rightarrow -\left(
15;15\right) ,$ $\,\,\,\,\,\left( 11;15\right) \rightarrow -\left(
11;15\right) ,\,\,\,\,\,\,\,\left( 15;11\right) \rightarrow +$ $\left(
15;11\right) $}

\vspace{0.25cm}

\item[D)]  {\small $\,\left( 15;15\right) $ $\rightarrow -\left(
15;15\right) ,$ $\,\,\,\,\,\left( 11;15\right) \rightarrow +\left(
11;15\right) ,\,\,\,\,\,\,\,\left( 15;11\right) \rightarrow -$$\left(
15;11\right) $$.$}
\end{enumerate}

\noindent It worth noting that for the construction of the non-chiral fusion
rules the first two construction bases along with the definition itself of
these rules was not taken explicitly in Ref.\cite{John1}. This has as a
consequence a great problem as we will see hereafter.

The consistency with the $\left( Z_{2}\right) $ action (A)$-$(D) leads each
one of its own to different structure of non-chiral fusion rules. To select
the physical structure the consistency of each one structure with the
bootstrap equations was taken into consideration. Therefore, by a counter
example the fusion rules constructed on the basis of the case (C) and (D)
are found inconsistent. The case (B) is \emph{presumed inconsistent} and
only the case (A) is retained.

\noindent The most striking result of the calculus of the structure
constants in the case (A) is the vanishing of one of the constants \cite
{John1}: 
\begin{equation}
C_{+++}=0  \label{spes}
\end{equation}
(with the notation $+=\left( 1,3\mid 13\right) ^{+}$) although the coupling $%
\left( +++\right) $ is permitted by the fusion rules. This result was
considered as specific by noting that `` $\cdots $ \textit{the vanishing \
of }$C_{+++}$ \textit{follows from our calculations and not from the fusion
rules}''. If one returns to the definition of the non-chiral fusion rules as
describing the operator content of the operator algebra the result (\ref
{spes}) found in Ref.\cite{John1} has nothing specific. In other words, this
result is simply not consistent with the bootstrap equations. The action of
the $Z_{2}$ symmetry that produce non-chiral fusion rules consistent with
the bootstrap equations is limited only to the case (B) which was not
considered in Ref.\cite{John1}. \linebreak In this case as an example of the
non-chiral fusion rules for the $D_{5}$ model we find: 
\begin{eqnarray}
\Phi ^{+}\cdot \Phi ^{+} &=&\left( 1,1\mid 1,1\right) +\Phi ^{+}+\left(
1,5\mid 1,5\right)  \nonumber \\
\Phi ^{-}\cdot \Phi ^{-} &=&\left( 1,1\mid 1,1\right) +\Phi ^{+}+\left(
1,5\mid 1,5\right)  \nonumber \\
\Phi ^{-}\cdot \Phi ^{+} &=&\Phi ^{-}  \label{ExempleF}
\end{eqnarray}
$\Phi ^{\pm }$ are the two copies of the degenerate field $\left( 1,3\mid
1,3\right) $.

The construction of $Z_{2}$ symmetry for the thermic subalgebra of the $D_{5}
$ model can be straighforwardly extended to the remaining fields of the $%
D_{5}$ model. In fact, the consistency of the $Z_{2}$ symmetry action with
the (OPA):

\medskip

\begin{tabular}{ll}
$\left( 1,1\mid 1,5\right) \times \left( 2,1\mid 2,1\right) =\left( 2,1\mid
2,5\right) ,$ & $\left( 1,1\mid 1,5\right) \times \left( 2,1\mid 2,1\right)
=\left( 2,5\mid 2,1\right) $ \\ 
$\left( 1,5\mid 1,5\right) \times \left( 2,1\mid 2,1\right) =\left( 2,5\mid
2,5\right) $ & 
\end{tabular}
\noindent

\medskip

\noindent leads to the following tree possibilities for the $Z_{2}$ action:

\begin{enumerate}
\item[B1)]  {\small $\left( 21;21\right) \rightarrow +\left( 21;21\right)
,\,\,\,\left( 25;25\right) \rightarrow +\left( 25;25\right) ,\,\,\,\left(
25;2,1\right) \rightarrow +\left( 25;21\right) ,\,\,\,\left( 21;25\right)
\rightarrow +\left( 21;25\right) .$}

\vspace{0.25cm}

\item[B2)]  {\small $\left( 21;21\right) \rightarrow -\left( 21;21\right)
,\,\,\,\left( 25;25\right) \rightarrow -\left( 25;25\right) ,\,\,\,\left(
25;2,1\right) \rightarrow +\left( 25;21\right) ,\,\,\,\left( 21;25\right)
\rightarrow +\left( 21;25\right) .$}

\vspace{0.25cm}

\item[B3)]  {\small $\left( 21;21\right) \rightarrow +\left( 21;21\right)
,\,\,\,\left( 25;25\right) \rightarrow +\left( 25;25\right) ,\,\,\,\left(
25;2,1\right) \rightarrow -\left( 25;21\right) ,\,\,\,\left( 21;25\right)
\rightarrow -\left( 21;25\right) .$}
\end{enumerate}

\noindent By a strainghtforward manipulation of the bootstrap equations in
these three cases, one can see that only the case (B3) is consistent with
bootstrap constraint. 

A first looking on the $Z_{2}$ symmetry for the $D_{5}$ model which we have
just determined, one remark that scalar fields are singlet under $Z_{2}$
contrary to spin fields which has a negative parity under $Z_{2}$. The
negative parity component $\left( \Phi ^{-}\right) $ of a degenerate field
behave like a spin field in the (OPA) (null spin structure field).

This construction of the $Z_{2}$ symmetry is done for the $D_{5}$ model
thermic subalgebra and is worth to be generalized for the remaining $D_{m}$
models, $m=1%
\bmod%
%
4$. Applying the same method as for the $D_{5}$ for the whole set of $D_{m}$ 
$\ \left( m=1%
\bmod%
%
4\right) $ is somewhat a delicate thing. The set of fields to take in
consideration increases with $\left( m\right) $ and the number of the \
possibilities of the $Z_{2}$ becomes, therefore, important. To overcome this
problem we shall consider a simple intuitive analysis that lays on the
following consideration: seeing that the symmetry $Z_{2}$ is introduced by
an \textit{ad-hoc} manner in order to separate the contribution of fields
that double in the (OPA) and having in mind the structure of the operator
content that is the same \footnote{%
This general symmetry structure is perceptible from the simple currents
construction of the $D$ series \cite{Amar}.} one is tempted to find that the 
$Z_{2}$ action follows a general law independently from the model (i.e. from 
$\left( m\right) $).

\noindent Thus the $Z_{2}$ symmetry action for the general $D_{m}$ models $%
\left( m=1%
\bmod%
%
4\right) $ have the same structure action as the $D_{5}$ model. {\ 
\begin{equation}
\,\,\,Z_{2}\left( \Phi ^{\pm }\right) =\pm \Phi ^{\pm
},\,\,\,\,\,\,\,\,Z_{2}\left( \Phi _{\alpha }^{s}\right) =-\Phi _{\alpha
}^{s},\,\,\,\,\,\,\,Z_{2}\left( \Phi _{\alpha }^{c}\right) =+\Phi _{\alpha
}^{c}.  \label{charge}
\end{equation}
} where $s=$ spin fields and $c=$ scalar fields.

Now as the action of the $Z_{2}$ symmetry is found the non-chiral fusion
rules are constructed by imposing the consistency of these rules with the
action of this symmetry. In other words this turn out to consider the
conservation of a parity charge in these rules. In consequence, the
non-chiral fusion rules in the $D_{m}$ case $m=1,2%
\bmod%
%
4,$ will have a $Z_{2}-grading$ structure at the same title as the cases $%
m=0,3%
\bmod%
%
4.$

\paragraph{Conclusion for the non-chiral fusion rules}

If we designate by $A_{0},$ the set of scalar fields and by $A_{1}$ the set
of spin and null spin fields the fusion rules will be of the forme 
\begin{equation}
A_{i}\ast A_{j}=A_{k},\,\,\,k=\left( i+j\right) 
\bmod%
%
2.  \label{low}
\end{equation}

\pagebreak

\section{Structure constants}

Now as the fusion rules are constructed it is possible then to solve the
bootstrap equations in order to obtain the structure constants. We present
thereafter the complete demonstration of this calculation. Since the $D_{m}$
models have the same non-chiral fusion rules structure (\ref{low}) we
restrict ourselves to the cases $m=1%
\bmod%
%
4$.

\subsection{Notations}

Here we sketch the most intriguing properties of conformal blocs in coulomb
gas construction (for more details see \cite{Dot} and \cite{Petko}). For
simplicity we limit ourselves to thermic subalgebra. First let us consider
the four point correlation functions: 
\begin{equation}
G\left( z,\overline{z}\right) =\left\langle \Phi _{N}\left( z_{1},\overline{z%
}_{1}\right) \cdot \Phi _{K}\left( z_{2},\overline{z}_{2}\right) \cdot \Phi
_{K}\left( z_{3},\overline{z}_{3}\right) \cdot \Phi _{N}\left( z_{4},%
\overline{z}_{4}\right) \right\rangle .  \label{corref}
\end{equation}
with $N\mathbf{=}\left( 1,n\mid 1,\overline{n}\right) $ denotes a field in
unitary minimal model spectrum and $z=\left( \frac{z_{12}z_{34}}{z_{13}z_{24}%
}\right) $. These correlations can be written in Coulomb gas formalism as
follows: 
\begin{equation}
G\left( z,\overline{z}\right) =f\left( z_{i}\right) f\left( \overline{z_{i}}%
\right) \sum_{i,j}^{n,\overline{n}}\gamma _{ij}\left( a,b,c\right)
I_{i}^{n}\left( a,b,c;z\right) \overline{I}_{j}^{n^{\prime }}\left( 
\overline{a},\overline{b},\overline{c};\overline{z}\right) .  \label{Decom}
\end{equation}
where $I_{i}\left( a,b,c;z\right) $ are conformal blocs in coulomb gas
formulation and $\gamma _{ij}\left( a,b,c\right) $ are coupling constants
with the notations:$\,\,a=2\alpha _{+}\alpha _{n},\,\,b=c=2\alpha _{+}\alpha
_{k}\,,d=2\alpha _{+}\left( 2\alpha _{0}-\alpha _{n}\right) \,\,\,\,\alpha
_{+}=\left[ \frac{m}{m+1}\right] ^{1/2},\alpha _{0}=-\left[ \frac{1}{m\left(
m+1\right) }\right] ^{1/2},\,\,\,\alpha _{i}=\frac{1}{2}\left( 1-i\right)
\alpha _{+}$ and $\,f\left( z_{i}\right) =\frac{\left( z_{14}\right)
^{2(h_{k}-h_{n})}}{\left( z_{13}z_{24}\right) ^{2h_{k}}}z^{2\alpha
_{n}\alpha _{m}}\left( 1-z\right) ^{2\alpha _{m}\alpha _{n}}.$

\noindent The conformal blocs have the short distance development:

{\ 
\begin{equation}
\lim_{z\rightarrow 0}I_{i}^{n}\left( a,b,c;z\right) =\left[ z^{-2\alpha
_{n}\alpha _{m}-\left( h_{n}+h_{k}-h_{p}\right) }\cdot \mathcal{N}%
_{i}^{n}\left( a,b,c\right) \cdot \left( 1+O\left( z\right) \right) \right]
_{p=n-k+2i-1}.  \label{Short}
\end{equation}
} $\mathcal{N}_{i}^{n}\left( a,b,c\right) $ is a normalization constant.

\noindent The correlations (\ref{Decom}) are in the t-channel development.
To express the s-channel development we use the conformal blocs
transformation under duality: 
\begin{equation}
I_{i}^{n}\left( a,b,c;z\right) =\sum_{j}\alpha _{ij}\left( a,b,c\right)
I_{j}^{n}\left( b,a,c;1-z\right) .  \label{duality}
\end{equation}
where $\alpha _{ij}\left( a,b,c\right) $ are elements of the monodromy
matrices. Therefore, the s-channel correlations can be written as follows: 
\begin{eqnarray}
G\left( z,\overline{z}\right) &=&f\left( z_{i}\right) \overline{f}\left( 
\overline{z_{i}}\right) \sum_{k,i,j,l}^{n,\overline{n}}\gamma _{ij}\left(
a,b,c\right) \alpha _{ik}\alpha _{jl}I_{k}^{n}\left( b,a,c;1-z\right) 
\overline{I}_{l}^{\overline{n}}\left( \overline{b},\overline{a},\overline{c}%
;1-\overline{z}\right) ,  \nonumber \\
&=&\sum_{i,j}^{n,\overline{n}}\gamma _{ij}\left( b,a,c\right)
I_{i}^{n}\left( b,a,c;1-z\right) \overline{I}_{j}^{\overline{n}}\left( 
\overline{b},\overline{a},\overline{c};1-\overline{z}\right) .
\label{Decom1}
\end{eqnarray}

\subsection{Bootstrap equations resolution}

Before solving the bootstrap equations for the $D_{m}$ series models let us
consider first this resolution for the simplest cases of $A_{m}$ series
models. For these models and from the (OPA) we can write the correlation
functions (\ref{corref}) at short distance as: 
\begin{equation}
G\left( z\right) \sim \sum_{P}\frac{\left( C_{N\,K}^{P}\right) ^{2}}{\left|
z\right| ^{2\left( h_{n}+h_{k}-h_{p}\right) }}\left( 1+O\left( z\right)
\right) .  \label{OPAA}
\end{equation}
where $\left( P\right) $ denotes a field permitted by the non-chiral fusion
rules law (\ref{FusionA}) of $\left( N\right) $ and $\left( K\right) $. The
consistency of the correlations in the s-channel with the non-chiral fusion
rules imposes that only the diagonal terms are present in (\ref{Decom}) and
that from (\ref{Short}) and (\ref{OPAA}) the structure constants are given
by: 
\begin{equation}
\left( C_{N\ K}^{P}\right) ^{2}\propto \left( \gamma _{i}\left( a,b,c\right) 
\mathcal{N}_{i}^{2}\left( a,b,c\right) \right) _{p=n-k+2i-1}.  \label{D1}
\end{equation}
The $\gamma _{i}\left( a,b,c\right) $ are obtained by imposing the
consistency of the correlations in the t-channel with non-chiral fusion
rules and which leads to the known result of (DF) namely: 
\begin{equation}
\sum_{k}\gamma _{k}\left( a,b,c\right) \alpha _{ki}\left( a,b,c\right)
\alpha _{kj}\left( a,b,c\right) =\gamma _{i}\left( b,a,c\right) \delta _{ij}.
\label{algeb}
\end{equation}

\noindent By imposing that $C_{N\,N}^{1}=1$ one can deduce the
proportionality factor in (\ref{D1}) and thus obtain the final forms of the
structure constants of the $\left( A\right) $ series models: 
\begin{equation}
C_{N\ K}^{P}=\sqrt{\frac{\gamma _{i}\left( a,b,c\right) }{\gamma _{1}\left(
b,a,c\right) }\,}\,\,\,\,\frac{\mathcal{N}_{i}\left( a,b,c\right) }{\mathcal{%
N}_{1}\left( b,a,c\right) }.  \label{D2}
\end{equation}

Now we propose to solve in the same manner the bootstrap equations for the $%
D_{m}$ series models. The correlations to deal with are of the form (\ref
{corref}) where $\mathbf{N}$ (with bold character) is a spin field $\left( 
\overline{n}=m+1-n\right) $ or a negative parity copies $\Phi ^{-}$ of a
degenerate field and $K$ is a scalar field $\left( \overline{m}=m\right) $.

\smallskip

\textbf{t-channel: }

In the channel-t the correlations at short distance written on the basis of
fusion rules (\ref{low}) are of the form: 
\begin{equation}
G_{1}=\sum_{p,\overline{p}}\frac{\left( C_{\mathbf{N}\,\,K}^{\mathbf{P}%
}\,\,\,\right) ^{2}\,\,\,\delta _{\overline{p},m+1-p}}{%
z_{12}^{h_{n}+h_{k}-h_{p}}z_{34}^{h_{n}+h_{k}-h_{p}}\overline{z}_{12}^{%
\overline{h}_{n}+\overline{h}_{k}-\overline{h}_{p}}\overline{z}_{34}^{%
\overline{h}_{n}+\overline{h}_{k}-\overline{h}_{p}}z_{24}^{2h_{p}}\overline{z%
}_{24}^{2\overline{h}_{p}}}.  \label{demo-t}
\end{equation}
In terms of the conformal blocs these correlations are written as: 
\begin{equation}
G_{1}=f\left( z_{i}\right) \overline{f}\left( \overline{z}_{i}\right)
\sum_{i,j}\gamma _{ij}^{\left( D\right) }\left( a,b,c\mid \overline{a}%
,b,c\right) I_{i}^{n}\left( a,b,c;z\right) I_{j}^{n}\left( \overline{a},b,c;%
\overline{z}\right) .  \label{demo-com}
\end{equation}
with: 
\begin{eqnarray*}
a &=&2\alpha _{+}\alpha _{n},\,\,\,\,\,\,\,\,\,\,\overline{a}=2\alpha
_{+}\alpha _{m+1-n}, \\
b &=&c=2\alpha _{+}\alpha _{k}.
\end{eqnarray*}
At short distance one has: 
\begin{eqnarray}
I_{i}\left( a,b,c;z\right) &\rightarrow &z^{-2\alpha _{k}\alpha _{n}-\left(
h_{n}+h_{k}-h_{p}\right) },  \nonumber \\
I_{j}\left( \overline{a},b,c;z\right) &\rightarrow &z^{-2\alpha _{\overline{n%
}}\alpha _{k}-\left( h_{\overline{n}}+h_{k}-h_{\overline{p}}\right) }.
\end{eqnarray}
with: 
\begin{equation}
\left\{ 
\begin{array}{c}
p=n-k+2i-1 \\ 
\overline{p}=\overline{n}-k+2j-1
\end{array}
\right. .  \label{demo-sys}
\end{equation}
For the combination of conformal blocs (\ref{demo-com}) to be consistent
with the non-chiral fusion rules expressed by the development (\ref{demo-t})
it is necessary that: $\overline{p}=m+1-p.$ If one takes this result in the
system (\ref{demo-sys}), along with the fact that $\overline{n}=m+1-n,$ one
finds that: $j=n+1-i.$

In consequence the consistency of the non-chiral fusion rules and the
combinations of conformal blocs at short distance at t-channel imposes that
only the coefficients $\gamma _{i,n+1-i}$ are non zero. Thus \footnote{%
Here we note the net difference between our form of correlation function in
the t-channel (\ref{demo-comt}) and the analogue (A.6) in the work \cite
{Petko}. One of these differences is the absence of signs factor in our
form. It is in order to lift this signs factor that the normalization of the
two point correlation functions was redefined in \cite{Petko}.}: 
\begin{equation}
G_{1}=f\left( z_{i}\right) \overline{f}\left( \overline{z}_{i}\right)
\sum_{i}\gamma _{i,n+1-i}^{\left( D\right) }\left( a,b,c\mid \overline{a}%
,b,c\right) I_{i}^{n}\left( a,b,c;z\right) I_{n+1-i}^{n}\left( \overline{a}%
,b,c;\overline{z}\right) .  \label{demo-comt}
\end{equation}
For more convenience, we adopt the following notation for the coupling
constants: 
\begin{equation}
\gamma _{i,n+1-i}^{\left( D\right) }\left( a,b,c\mid \overline{a},b,c\right)
=\gamma _{i}^{\left( D\right) }\left( a,b,c\right) =\gamma _{n+1-i}^{\left(
D\right) }\left( \overline{a},b,c\right) .
\end{equation}
The structure constants are obtained as limits at short distance of (\ref
{demo-comt}) 
\begin{equation}
\left( C_{\mathbf{N}\,\,K}^{\mathbf{P}}\,\,\,\right) ^{2}\,\propto \gamma
_{i}^{\left( D\right) }\left( a,b,c\right) \mathcal{N}_{i}\left(
a,b,c\right) \mathcal{N}_{n+1-i}\left( \overline{a},b,c\right) .
\label{cont-D}
\end{equation}

\textbf{s-channel: }

In order to impose the bootstrap constraint we will develope (\ref{demo-comt}%
) in the s-channel. This is done by considering the duality transformation
of the conformal blocs (\ref{duality}). Thus, the s-channel correlation
functions are of the form: 
\begin{eqnarray}
G_{1} &=&f\left( z_{i}\right) \overline{f}\left( \overline{z}_{i}\right)
\sum_{i,l,l^{\prime }}\gamma _{i}^{\left( D\right) }\left( a,b,c\right)
\,\alpha _{il}\left( a,b,c\right) \alpha _{n+1-i,l^{\prime }}\left( 
\overline{a},b,c\right) \,  \nonumber \\
&&\,\,\,\,\,\,\,\,\,\,\,\,\,\,\,\,\,\,\,\,\,\,\,\,\,\,\,\,\,\,\,\,\,\,\,\,\,%
\,\,\,\,\,\,\,\,\,\,\,\,\,\,\,\,\,\,\,\,\,\,\,\,\,\,\,\,I_{l}^{n}\left(
b,a,c;1-z\right) \,\,I_{l^{\prime }}^{n}\left( b,\overline{a},c;1-\overline{z%
}\right) .  \label{Forms}
\end{eqnarray}
This form of the correlations must be consistent at short distance with the
non-chiral fusion rules which state that only scalar fields are present in
the fusion of two fields of the same nature. In other words only the
diagonal terms are present in the correlations (\ref{Forms}): 
\begin{equation}
G_{1}=f\left( z_{i}\right) \overline{f}\left( \overline{z}_{i}\right)
\sum_{l}\gamma _{l}^{\left( D\right) }\left( b,a,c\right)
\,I\,_{l}^{n}\left( b,a,c;1-z\right) \,\,I_{l^{\prime }}^{n}\left( b,%
\overline{a},c;1-\overline{z}\right) .  \label{Forms0}
\end{equation}
By comparing these two latter forms of s-channel correlations one can deduce
that the coupling constants $\gamma _{i}^{\left( D\right) }\left(
a,b,c\right) $ are solutions of the algebraic equation: 
\begin{equation}
\sum_{i}\gamma _{i}^{\left( D\right) }\left( a,b,c\right) \,\alpha
_{il}\left( a,b,c\right) \alpha _{n+1-i,l^{\prime }}\left( \overline{a}%
,b,c\right) =\gamma _{l}^{\left( D\right) }\left( b,a,c\right) \delta
_{ll^{\prime }}.  \label{algD}
\end{equation}
The problem of determining the structure constants of the $\left(
D_{m}\right) $ series models is reduced then to the resolution of the
algebraic equation (\ref{algD}). For this goal, we consider the following
analytic property of conformal blocs \cite{Petko}: 
\begin{equation}
I_{i}^{n}\left( a,b,c;z\right) =z^{-2\alpha _{n}\alpha _{m}-\left(
h_{n}+h_{m}-h_{p}\right) }I_{n+1-i}\left( d,c,b;z\right) .  \label{demo-chan}
\end{equation}
This will permit in fact to deduce that: 
\begin{equation}
\alpha _{n+1-i,l^{\prime }}\left( \overline{a},b,c\right) =\alpha
_{i,l^{\prime }}\left( \overline{d},c,b\right) .
\end{equation}
with 
\[
\overline{d}=2\alpha _{+}\left( 2\alpha _{0}-\alpha _{m+1-k}\right) . 
\]
At this level we were inspired by Petkova's work \cite{Petko}; i.e. by using
the fact that: 
\begin{equation}
\overline{d}-a=\left( 2-m\right) \in \Bbb{N}.  \label{prop1}
\end{equation}
and that in these conditions: 
\begin{equation}
\alpha _{il}\left( \overline{d},c,b\right) =\alpha _{il}\left( \overline{d}%
,b,c\right) =\left( -1\right) ^{\left( \overline{d}-a\right) \left(
l-1\right) }\alpha _{il}\left( a,b,c\right) .  \label{prop2}
\end{equation}
Now if we report these relations in (\ref{algD}) we arrive at the equation: 
\begin{equation}
\sum_{i}\gamma _{i}^{\left( D\right) }\left( a,b,c\right) \,\alpha
_{il}\left( a,b,c\right) \alpha _{il^{\prime }}\left( a,c,b\right) =\left(
-1\right) ^{\left( m-2\right) \left( l-1\right) }\gamma _{l}^{\left(
D\right) }\left( b,a,c\right) \delta _{ll^{\prime }}.  \label{algD1}
\end{equation}
To find the solutions of this equation we consider its analogue of the $%
\left( A\right) $ series (\ref{algeb}). By comparison we can derive the
solutions of (\ref{algD1}) under the form: 
\begin{eqnarray}
\gamma _{i}^{D}\left( a,b,c;z\right) &=&\gamma _{i}^{A}\left( a,b,c;z\right)
,  \label{resu1} \\
\gamma _{i}^{D}\left( b,a,c;z\right) &=&\left( -1\right) ^{\left( m-2\right)
\left( i-1\right) }\gamma _{i}^{A}\left( b,a,c;z\right) .  \label{resu2}
\end{eqnarray}
Once the coupling constants are determined the structure constants of the$%
\left( D_{m}\right) $ series models are given by: 
\begin{equation}
\left( C_{\mathbf{N}\,\,K}^{\mathbf{P}}\,\,\,\right) ^{2}\,\propto \gamma
_{i}^{\left( A\right) }\left( a,b,c\right) \mathcal{N}_{i}\left(
a,b,c\right) \mathcal{N}_{n+1-i}\left( \overline{a},b,c\right) .
\label{consD1}
\end{equation}
To write this last result in more convenient form we use the fact that if $%
\left| a^{\prime }-a\right| $ is an integer then\footnote{%
This can be readely deduced from (\ref{algeb}) by using the properties (\ref
{prop1}) and (\ref{prop2}).}: 
\[
\gamma _{i}^{\left( A\right) }\left( a,b,c\right) =\gamma _{i}^{\left(
A\right) }\left( a^{\prime },b,c\right) . 
\]
and immediately we can show that: 
\[
\gamma _{i}^{\left( A\right) }\left( a,b,c\right) =\gamma _{n+1-i}^{\left(
A\right) }\left( d,c,b\right) =\gamma _{n+1-i}^{\left( A\right) }\left( 
\overline{a},c,b\right) =\gamma _{n+1-i}^{\left( A\right) }\left( \overline{a%
},b,c\right) . 
\]
Using this last result in (\ref{consD1}) we find finally: 
\begin{eqnarray}
\left( C_{\mathbf{N}\,\,K}^{\mathbf{P}}\,\,\,\right) ^{2} &\propto &\left( 
\sqrt{\gamma _{i}^{\left( A\right) }\left( a,b,c\right) }\mathcal{N}%
_{i}\left( a,b,c\right) \right) \left( \sqrt{\gamma _{n+1-i}^{\left(
A\right) }\left( \overline{a},b,c\right) }\mathcal{N}_{n+1-i}\left( 
\overline{a},b,c\right) \right) ,  \nonumber \\
&=&C_{N\,\,K}^{P}\cdot C_{\overline{N\,}\,K}^{\overline{P}}\,\,\,.
\label{constD2}
\end{eqnarray}
We see thus that the structure constants of the $D_{m}$ series factorizes
out in those of the chiral algebra expressed by the $A_{m}$ series. Another
important result can be deduced from (\ref{resu2}) concerning the signs of
the product of the structure constants namely: 
\[
S\left( C_{\mathbf{N\,\,N}}^{F}D_{K\,K}^{F}\right) =\left( -1\right)
^{\left( m-2\right) \left( \frac{F-1}{2}\right) }. 
\]
Since the scalar fields constitute a subalgebra in the (OPA) we can chose
the signs of the structure constants of this subalgebra arbitrarily. By
opting for positive signs we can deduce that: 
\begin{equation}
S\left( C_{\mathbf{N\,\,N}}^{F}\right) =\left( -1\right) ^{\left( m-2\right)
\left( \frac{F-1}{2}\right) }  \label{signes}
\end{equation}
This last result is obtained by the resolution of the bootstrap equations
realized from duality symmetry of correlations of the form (\ref{corref}).
To determine the signs of the structure constants between general couplings $%
C_{\mathbf{N\,\,K}}^{F}$ we must consider more general correlation forms.
This can be readily done by considering the general properties \linebreak
(which generalize (\ref{prop2})) of the breading (monodromy) matrices
determined by the connection established in Refs.\cite{Petko, Zuber}. We
find that the result (\ref{signes}) is general . 
\begin{equation}
S\left( C_{\mathbf{N\,\,K}}^{F}\right) =\left( -1\right) ^{\left( m-2\right)
\left( \frac{F-1}{2}\right) }  \label{signe}
\end{equation}

\section{Discrete symmetries of minimal models}

We make appear a $Z_{2}$ discrete symmetry in the construction of the
non-chiral fusion rules. This symmetry has appeared automatically in the
automorphism invariant cases through the $Z_{2}-grading$ structure of the
non-chiral fusion rules. In the integer invariant cases, the $Z_{2}$
symmetry was put into evidence in a different manner and this as a
consequence of the existence of two copies of some scalar fields. The fusion
rules obtained have also a $Z_{2}-grading$ structure.

Essentially, the $Z_{2}$ symmetry appear for the whole set of the $D_{m}$
series as a consequence of the $Z_{2}-grading$ structure of the fusion
rules. These structure expresses the conservation of a parity charge in
these rules. In fact, the $Z_{2}$ permit to associate a positive parity to
the scalar fields and a negative one to spin and null spin fields. This
important fact suggests us to find a physical interpretation of the $\left(
Z_{2}\right) $ symmetry.

\subsection{The $ADE$ classification as lattice models}

The fact that the universal critical properties are controlled by the long
range fluctuations enable to treat them by a continuum field theory;
conformal invariant at the critical point. The richness of the conformal
symmetry in two dimensions makes it possible the classification of the
universality classes. The $ADE$ classification of the minimal models present
a typical example of such classification.

In addition the universality principle of critical phenomena makes it
conceivable to construct a statistical model of spin on lattice for all
universality classes(conformal model). Thus, one finds that the critical
properties of the critical and the tricritical three states Potts models are
given by the $D$ series with $m=5$ and $m=4$ respectively; the Ising model
at the other hand is described by the $A$ series with $m=3$.

Nowadays it is established that the whole set of unitary minimal models of
the $ADE$ classification expresses the critical proprieties of the models
said (RSOS)\textbf{\ }\cite{RSOS}. The formulation of these models (RSOS) is
realized on the basis of the simple Lie algebra of type $ADE$ where at each
site of the lattice is attributed a \emph{weight} variable. These weights
correspond to those of Coxeter-Dynkin diagram for a simple Lie $ADE$ algebra
with the condition that two closest neighbors have neighboring weights in
the Coxeter-Dynkin diagram.

None of these models has a continuum symmetry but on the contrary they have
discrete symmetries. In this respect, we find as an example that the three
state Potts models have a discrete symmetry $S_{3}$ which is a sum of cyclic
discrete symmetries $Z_{3}$ and $Z_{2}$ and the Ising model has a $Z_{2}$
cyclic symmetry. For the whole set of RSOS models the discrete symmetries
are nothing but the automorphism group of the Coxeter-Dynkin diagrams. In
consequence the whole set of these models have a $Z_{2}$ symmetry except the 
$D_{4},D_{5},E_{7}$ and $E_{8}$ models. The diagrams of $D_{5}$ and $D_{4}$
have an $S_{3}$ symmetry and those of $E_{7}$ and $E_{8}$ have no symmetry.

The minimal models and their $ADE$ classification involve in their
construction only the conformal symmetry and modular invariance (i.e.
periodic) of partition functions. So it is of importance to know if this
classification is consistent with the presence of other discrete symmetries.
This is very important, indeed, because the critical properties have a
strong dependence on symmetries and if the $ADE$ classification describes
the critical behavior ( universality classes) of the RSOS models then it
must be consistent with the presence of discrete symmetries of these lattice
models. This is exactly what was done in a recent work \cite{Ruelle} where
the consistency of the $ADE$ classification with the presence of discrete
cyclic $\left( Z_{n}\right) $ symmetries is was investigated. The result
found therein confirm that only symmetries which are present in the RSOS
models are consistent with the $ADE$ classification. Another important
result determined in Ref.\cite{Ruelle} is the action of these discrete
symmetries on primary fields. For the particular cases of the $D$ series and 
$Z_{2}$ symmetry this action is exactly identical to that of our $Z_{2}$
symmetry, found in the construction of the non-chiral fusion rules. Instead
the discrete symmetry $Z_{2}$ of the $D$ series RSOS models appear through 
\emph{the consistency of the non-chiral fusion rules with the action of this
symmetry}.

This important constatation leads us to think to construct for the three
state Potts models $D_{5}$ $\left( D_{4}\right) $ the non-chiral fusion
rules consistent with the other symmetry of these models namely $\left(
Z_{3}\right) $ symmetry. This construction will be done for the $D_{5}$
critical three state Potts model and it will be available by symmetry to the 
$D_{4}$ tricritical three state Potts model.

\subsection{$Z_{3}$ construction for $D_{5}$ model}

The critical three states Potts model is a spin lattice model with discrete
spin complex variable $\sigma =\exp \left( i\phi \right) ;\,\,\phi =0,\pm 
\frac{2\pi }{3}.$ The lattice Hamiltonian of this model is given by: 
\begin{equation}
H=J\sum_{x,i}\frac{1}{2}\left( \sigma _{x}\sigma _{x+i}^{*}+\sigma
_{x}^{*}\sigma _{x+i}\right)  \label{Hamil}
\end{equation}
where $\left( x\right) $ denotes lattice spin position and $\left( i\right) $
the neighboring position.

At the scaling limit the discrete spin variables become continues operators $%
\sigma \left( x\right) .$ It is natural also to identify the density energy
operator $\varepsilon \left( x\right) $ from (\ref{Hamil}) as the scaling
limit of the interaction term $\sigma _{x}\sigma _{x+i}^{*}+\sigma
_{x}^{*}\sigma _{x+i}$. These two identified operator were known to have a
scaling dimension equal to $\Delta _{\sigma }=\frac{2}{15}$ for the spin
complex operator $\sigma \left( x\right) \left( \sigma ^{*}\left( x\right)
\right) $ and $\Delta _{\varepsilon }=\frac{4}{5}$ for the energy density
operator. The complex nature of the spin variable $\sigma \left( x\right) $
is one of the reasons that the critical three states Potts model is
identified with the $D_{5}$ model rather than the diagonal $A_{5}$ model 
\cite{John2}. In fact, from the spin variable we can define two real spin
variables: $\sigma +\sigma ^{*}$ and $\sigma -\sigma ^{*}$ which reflect the
presence of two copies of the same scalar real primary field in the model.
As a consequence and from the operator content of the $D_{5}$ we can
identify $\sigma +\sigma ^{*}$ and $\sigma -\sigma ^{*}$ with the two copies 
$\Phi ^{\pm }$ of \ the doubled field $\left( 2,3\mid 2,3\right) $ and hence
write the complex spin variable as: 
\begin{equation}
\sigma =\frac{1}{\sqrt{2}}\left( \Phi ^{+}+i\Phi ^{-}\right)  \label{spinv}
\end{equation}
In addition the energy density field can be identified with the scalar field 
$\left( 2,1\mid 2,1\right) $.

The Hamiltonian (\ref{Hamil}) is invariant under the discrete cyclic
symmetry $\left( Z_{3}\right) $ defined with its action on the spin variable
as: 
\begin{equation}
Z_{3}\left( \sigma \left( x\right) \right) =\exp \left( \frac{2\pi i}{3}%
\right) \sigma \left( x\right)  \label{Z3}
\end{equation}

\noindent From the operator product expansion, one can show that the second
doubled field $\left( 2,3\mid 2,3\right) $ which can be represented as a
complex field $\left( \Omega \right) $ like (\ref{spinv}) transforms under $%
\left( Z_{3}\right) $ in the same way as $\sigma \left( x\right) $ (\ref{Z3}%
). The other non doubled fields were invariant under $\left( Z_{3}\right) $
because each of them are conjugate\footnote{%
The conjugate field $\Phi ^{c}$ of a field $\Phi $ is defined by the
condition that the OPA of these two fields produce the identity field: 
\[
\Phi ^{c}\cdot \Phi \sim 1 
\]
} to itself in the operator product expansion.

\subsubsection{The $Z_{3}$ non-chiral fusion rules}

From the action of the $Z_{3}$ symmetry we can easily deduce, as was done
for the $Z_{2}$ symmetry, the non-chiral fusion rules consistent with its
action. This is what was considered in a second work \cite{John2} by the
author of Ref.\cite{John1}. For example one finds that: 
\begin{eqnarray}
\Omega \times \Omega &=&\Omega ^{\ast }  \nonumber \\
\Omega \times \Omega ^{\ast } &=&1+\left( 1,5\mid 1,5\right) +\left( 1,5\mid
11\right) +\left( 1,1\mid 1,5\right)  \nonumber \\
\Omega \times \left( 1,1\mid 1,5\right) &=&\Omega  \nonumber \\
\Omega \times \left( 2,1\mid 2,1\right) &=&\sigma +\Omega  \label{Z3rules}
\end{eqnarray}
In \cite{John2} it was noted that the solutions obtained from the two
constructions namely $\left( Z_{2}\right) $ and $\left( Z_{3}\right) $ are 
\textit{``..., of course, inequivalent.''. }If the two constructions are
really inequivalent one can deduce that if (\ref{Z3rules}) produce the (OPA)
content of the critical three states Potts model then the $\left(
Z_{2}\right) $ (OPA) structure describes another critical model, which is
ambiguous. The $\left( Z_{2}\right) $ as $\left( Z_{3}\right) $ are both
discrete symmetries of the three state Potts model so that the (OPA)
structure obtained from these two symmetries must be equivalent. The error
committed in \cite{John1} is that the good action (the (A) case rather than
the (B) case) of the $\left( Z_{2}\right) $ symmetry was not considered. We
propose now to establish the equivalence between the $\left( Z_{3}\right) $
and $\left( Z_{2}\right) $ structure of the (OPA). This is done easily by
calculating the fusion for example of $\Omega $ and $\Omega ^{\ast }$. From
the rules (\ref{ExempleF}) one finds that: 
\begin{eqnarray}
\Omega \times \Omega ^{\ast } &=&\frac{1}{2}\left( \Phi ^{+}+i\Phi
^{-}\right) \times \left( \Phi ^{+}-i\Phi ^{-}\right)  \nonumber \\
&=&\frac{1}{2}\left[ \Phi ^{+}\times \Phi ^{+}+\Phi ^{-}\times \Phi
^{-}+i\left( \Phi ^{+}\times \Phi ^{-}-\Phi ^{-}\times \Phi ^{+}\right) %
\right]  \nonumber \\
&\Leftrightarrow &\left. \left[ 
\begin{array}{c}
1+C_{+\,+\,+}+C_{+\,+\,\left( 15\mid 15\right)
}+1+C_{-\,-\,+}+C_{-\,-\,\left( 15\mid 15\right) } \\ 
+i\left( 
\begin{array}{c}
C_{+\,-\,-}+C_{+\,-\,\left( 15\mid 11\right) }+C_{+\,-\,\left( 11\mid
15\right) }-C_{-\,+\,-} \\ 
-C_{-\,\,+\,\left( 15\mid 11\right) }-C_{-\,\,+\,\left( 11\mid 15\right) }
\end{array}
\right)
\end{array}
\right] \right.
\end{eqnarray}
Using the fact that: $C_{-\,-\,+}=-C_{+\,+\,+}$ and $C_{+\,+\,\left(
15,15\right) }=C_{-\,-\,\left( 15,15\right) }$ deduced from the signs of the
structure constants (\ref{signe}) and the fact that: 
\[
C_{abc}=\left( -1\right) ^{s\left( a\right) +s\left( b\right) +s\left(
c\right) }C_{bac}=\left( -1\right) ^{s\left( a\right) +s\left( b\right)
+s\left( c\right) }C_{cba} 
\]
we deduce that: 
\[
\Omega \times \Omega ^{\ast }=1+\left( 1,5\mid 1,5\right) +\left( 1,5\mid
1,1\right) +\left( 1,1\mid 1,5\right) 
\]
In the same way we can proof that: 
\begin{equation}
\Omega \ast \Omega =\Omega ^{\ast }  \label{last}
\end{equation}
What we have just proved through equation (\ref{last}) is the equivalence of
the two constructions of the non-chiral fusion rules based on the $Z_{2}$
and $Z_{3}$ symmetries. What we have exactly done is the following: If we
consider the non-chiral fusion rules as a commutative and associative ring
(in the same way as the fusion rules \cite{Fuchs}) with as a basis the set
of primary fields of the $D_{m}$ model then by transformation (\ref{spinv})
we have achieved a change of basis. The $Z_{2}$ symmetry structure of the
fusion rules in the real basis is manifested by the complex $Z_{3}$ symmetry
in the new complex basis (\ref{spinv}).

The question that is of interest to answer at this stage is the very
prediction of the existence of the complex cyclic symmetry $\left(
Z_{N}\right) $ for the other models of the $\left( D_{m}\right) $ series as
we have done for the $D_{5}\left( D_{4}\right) $ models of the $\left(
Z_{3}\right) $ symmetry. Indeed if such symmetry exists it cannot be but a
cyclic symmetry of order $3$ $\left( Z_{3}\right) $. This turns out to prove
the equivalent of the equation (\ref{last}) with: 
\begin{eqnarray}
\Omega &=&\frac{1}{\sqrt{2}}\left( \Phi ^{+}+i\Phi ^{-}\right)  \label{chan}
\\
\Phi ^{\pm } &=&\left( 1,\frac{m+1}{2}\mid 1,\frac{m+1}{2}\right) ^{\pm } 
\nonumber
\end{eqnarray}
In fact, the change to the complex basis cannot be consistent with the (OPA)
i.e. $\Phi \times \Phi ^{*}\sim 1$; but for the values of $m=5\bmod8\left(
m=6\bmod8\right) $ \ and for this cases precisely and by following the same
approach as for $\left( D_{5}\right) $ model we can show that it is not
possible to have a form consistent with (\ref{last}). This meets the results
found in Ref.\cite{Ruelle}.

\section{General discussions}

In this work we have proposed an approach to solve the bootstrap equations
in the case of the minimal and unitary models of $D_{m}$ series. This
approach consists in the very construction of the non-chiral fusion rules
which determines the operator content of the operator algebra. Once these
fusion rules are determined it will be possible to solve the bootstrap
equations by considering the consistency of these equations at short
distance with these rules.

For the $D_{m}$ series models the non-chiral fusion rules found have a $%
Z_{2}-grading$ structure. This later reflects, directly in the automorphism
invariant cases, the existence of a $Z_{2}$ symmetry. In the integer
invariant cases the $Z_{2}-grading$ structure was instead deduced as a
consequence of the doubling of certain scalar fields and therefore the
existence of discrete parity symmetry $Z_{2}$. This symmetry has been
interpreted as scaling limit of $Z_{2}$ symmetry of the $D$-like (RSOS) spin
lattice models. In addition, beginning from the non-chiral fusion rules
consistent with the $Z_{2}$ action and the signs of the structure constants,
we succeeded in finding the non-chiral fusion rules consistent with $Z_{3}$
symmetry for the three state Potts models $D_{5}\left( D_{4}\right) $. Also
we have proved that the existence of other discrete cyclic symmetries was
not possible for the remaining of the $D_{m}$ models. Regarding the RSOS
lattice construction this symmetries structure of the $D_{m}$ model is
nothing but the automorphism group of the $D$ Coxeter-Dynkin diagrams \cite
{RSOS}.

A Further important interpretation of the manifestation of these discrete
symmetries for minimal models is given in \cite{Kastor}. In fact, Goddard,
Kent and Olive (GKO) \cite{GKO} gave a ``coset'' construction to generate
the unitary minimal models from unitary representation of $SU_{2}\left(
K\right) \times SU_{2}\left( 1\right) /SU_{2}\left( K+1\right) $. The $%
SU_{2} $ models are $Z_{2}$ invariant and so is the (GKO) coset construction
of minimal unitary models. At another hand, it turns out that the three
state Potts models can also be realized as coset construction of $%
SU_{3}\left( K\right) \times SU_{3}\left( 1\right) /SU_{3}\left( K+1\right) $
which is $Z_{3}$ invariant. Thus these last models carry a $Z_{3}$ as well
as $Z_{2}$ symmetry.

These remarks give to our approach a possibility to be generalized to other
conformal models namely rational models. For these models and particularly
in the case of Kac-Moody chiral algebras the non-chiral fusion rules in the $%
D-$like series may be structured following a discrete symmetry which is a
center or a subalgebra of the center of the chiral algebra. For a coset
construction $g/h$ the discrete symmetry is a subalgebra of the center of $g$
that preserves $h$. A convenient manner to formulate the problem is
presented in another work of the present auctors \cite{Amar}. This later is
based, in the same way as in this work, on the consistency of the non-chiral
fusion rules with the chiral fusion rules and with the operator content
derived from the modular constraint. The discrete symmetry structure is
introduced simply by the simple currents construction of $D-$like series. \
In this framework this symmetry is nothing but the\emph{\ effective center}
of the simple currents utilized in modular invariant construction \cite
{Simple}.

Finally it is important to mention a possible connection of the discrete
symmetry structure of the non-chiral fusion rules with a reflection group of
what is known as graphs construction \cite{Zuber, Graphs}. These graphs are
a generalization of the $ADE$ Coxeter-Dynkin diagrams; so a possible
integrable lattice interpretation and construction of the (OPA) and
structure constants my be envisaged.

\begin{quote}
\textbf{Acknowledgments: }We are grateful to Prof. J. McCabe for the helpful
discussions we have had together and which were the cornerstone of the
present work. One of us A.R. expresses his thanks to H.Houili for reading
the manuscript.
\end{quote}

\end{document}